# Theoretical Analysis of Quartz Plate Acoustic Wave Resonators and Sensors Using Three-Dimensional Equations of Anisotropic Elasticity or Piezoelectricity

Jiashi Yang (jyang1@unl.edu)
Department of Mechanical and Materials Engineering
University of Nebraska-Lincoln, Lincoln, NE 68588-0526, USA

**Abstract** – This is a review of theoretical results from the three-dimensional equations of anisotropic elasticity or linear piezoelectricity on waves and vibrations in quartz crystal plates. It covers both the classical results on acoustic wave resonators and the relatively new applications in acoustic wave sensors. It discusses the basic thickness modes with various complications due to electrodes, mass loading, contact with fluids, and air gaps, etc. as well as the more complicated transversely varying modes. These results are fundamental for the understanding and design optimization of the widely used quartz crystal plate resonators and sensors.

## I. INTRODUCTION

Piezoelectric crystals exhibit electromechanical coupling. They experience mechanical deformations when placed in an electric field, and become electrically polarized under mechanical loads. These materials have been used for a long time to make various electromechanical devices. Examples include transducers for converting electrical energy to mechanical energy or vice versa, and acoustic wave resonators providing frequency standards for time keeping and frequency operation. Resonator frequency shifts caused by various effects like a temperature change, stress, surface additional mass or contact with a fluid are the foundation of resonator-based acoustic wave sensors.

Quartz is probably the most widely used material for crystal resonators. Due to material anisotropy and piezoelectric coupling, theoretical analysis of quartz resonators using the three-dimensional (3-D) theories of anisotropic elasticity or piezoelectricity usually presents considerable mathematical challenges. Since a big portion of quartz resonators operate with the thickness-shear modes of a plate, researchers have developed approximate two-dimensional (2-D) theories to model vibrations of piezoelectric plates and employed these equations in the analysis and design of quartz resonators. The 2-D plate equations are obtained by expanding the electromechanical fields in the plate into power series of the plate thickness coordinate and taking moments of various orders about the plate middle plane. Usually the lower-order equations are used. Theoretical results on quartz resonators obtained from the approximate 2-D plate equations were reviewed systematically in [1]. However, the more fundamental theoretical results from the exact 3-D equations are still scattered in the literature. It will be convenient to researchers if they are gathered in one place.

In this paper we review theoretical results from the 3-D equations of linear elasticity or piezoelectricity on waves propagating in and vibrations of quartz crystal plates. Some of the results summarized were obtained from direct approximations of the 3-D equations by dropping some small terms. Both the classical results on quartz resonators and their relatively new applications in various acoustic wave sensors are included. After a brief presentation of the basics of a quartz plate in Section II, the review begins with the simplest thickness modes in Section III with consideration of various effects due to mass layers, contact with fluids, and air gaps, etc., and then proceeds to the more complicated shear-horizontal modes in Section IV and straight-crested modes varying along the digonal axis of rotated Y-cut quartz plates in Section V. Section VI is on modes varying with both of the in-plane coordinates. Finally, some conclusions are drawn in Section VII.

## II. QUARTZ PLATES

Quartz belongs to the trigonal crystal class of 32 or $D_3$ [2]-[4]. It has one axis of trigonal symmetry and, in the plane at right angles, three digonal axes 120º apart. Its independent elastic, piezoelectric, and dielectric constants are six, two, and two, respectively. Quartz plates taken out of a bulk crystal at different orientations are referred to as plates of different cuts. With respect to the crystallographic axes ($X,Y,Z$) where $Z$ is the trigonal axis and $X$ a digonal axis, a particular cut is specified by two angles, $\phi$ and $\theta$, with respect to ($X,Y,Z$) (see Fig. 1) and is called a doubly-rotated cut in general. For example, the stress-compensated SC-cut has ($\phi,\theta$) = (22.4°,-33.88°). Plates of different cuts have different material matrices with respect to the coordinate system ($x_1,x_2,x_3$) in and normal to the plane of the plates.

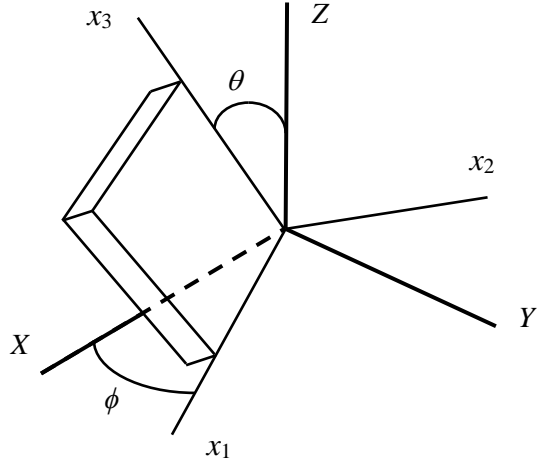

Fig. 1. A quartz plate cut from a bulk crystal.

One class of cuts of quartz plates, called rotated Y-cuts, have $\phi = 0$ and they are called singly-rotated cuts. They include the most widely used AT-cut with ($\phi,\theta$) = (0°,-35°) as a special case. Rotated Y-cut quartz plates exhibit monoclinic symmetry of class 2 or $C_2$ in the plate coordinate system ($x_1,x_2,x_3$), with $x_1$ a digonal axis. They are very useful in applications and most theoretical results for quartz plates obtained from the 3-D equations are for monoclinic crystals. Their material matrices and the governing equations for the displacement vector **u** and the electric potential $\varphi$ are [3, 4]

$$\begin{pmatrix} c_{11} & c_{12} & c_{13} & c_{14} & 0 & 0 \\ c_{21} & c_{22} & c_{23} & c_{24} & 0 & 0 \\ c_{31} & c_{32} & c_{33} & c_{34} & 0 & 0 \\ c_{41} & c_{42} & c_{43} & c_{44} & 0 & 0 \\ 0 & 0 & 0 & 0 & c_{55} & c_{56} \\ 0 & 0 & 0 & 0 & c_{65} & c_{66} \end{pmatrix}, \begin{pmatrix} e_{11} & e_{12} & e_{13} & e_{14} & 0 & 0 \\ 0 & 0 & 0 & 0 & e_{25} & e_{26} \\ 0 & 0 & 0 & 0 & e_{35} & e_{36} \end{pmatrix}, \begin{pmatrix} \varepsilon_{11} & 0 & 0 \\ 0 & \varepsilon_{22} & \varepsilon_{23} \\ 0 & \varepsilon_{32} & \varepsilon_{33} \end{pmatrix}, \quad (1)$$

$$\begin{aligned} & c_{11}u_{1,11} + (c_{12}+c_{66})u_{2,12} + (c_{13}+c_{55})u_{3,13} + (c_{14}+c_{56})u_{2,13} + (c_{14}+c_{56})u_{3,12} + 2c_{56}u_{1,23} + c_{66}u_{1,22} + c_{55}u_{1,33} \\ & \quad + e_{11}\varphi_{,11} + e_{26}\varphi_{,22} + (e_{36}+e_{25})\varphi_{,23} + e_{35}\varphi_{,33} = \rho\ddot{u}_1, \\ & c_{56}u_{3,11} + (c_{56}+c_{14})u_{1,13} + (c_{66}+c_{12})u_{1,12} + c_{66}u_{2,11} + c_{22}u_{2,22} + (c_{23}+c_{44})u_{3,23} + 2c_{24}u_{2,23} + c_{24}u_{3,22} \\ & \quad + c_{34}u_{3,33} + c_{44}u_{2,33} + (e_{26}+e_{12})\varphi_{,12} + (e_{36}+e_{14})\varphi_{,13} = \rho\ddot{u}_2, \\ & c_{55}u_{3,11} + (c_{55}+c_{13})u_{1,13} + (c_{56}+c_{14})u_{1,12} + c_{56}u_{2,11} + c_{24}u_{2,22} + 2c_{34}u_{3,23} + (c_{44}+c_{23})u_{2,23} + c_{44}u_{3,22} \\ & \quad + c_{33}u_{3,33} + c_{34}u_{2,33} + (e_{25}+e_{14})\varphi_{,12} + (e_{35}+e_{13})\varphi_{,13} = \rho\ddot{u}_3, \\ & e_{11}u_{1,11} + (e_{12}+e_{26})u_{2,12} + (e_{13}+e_{35})u_{3,13} + (e_{14}+e_{36})u_{2,13} + (e_{14}+e_{25})u_{3,12} + (e_{25}+e_{36})u_{1,23} + e_{26}u_{1,22} \\ & \quad + e_{35}u_{1,33} - \varepsilon_{11}\varphi_{,11} - \varepsilon_{22}\varphi_{,22} - 2\varepsilon_{23}\varphi_{,23} - \varepsilon_{33}\varphi_{,33} = 0. \end{aligned} \quad (2)$$

(2) is a very complicated system of partial differential equations. For finite plates the situation is further complicated by boundary conditions. Even with some approximations, analytical solutions to these equations are difficult to obtain.

Quartz has relatively weak piezoelectric coupling. For resonator and sensor applications, its relevant electromechanical coupling coefficient $k_{26}$ for the AT-cut is on the order of 1% [3, 4]. Therefore, for frequency analysis of a quartz plate, the small piezoelectric coupling can often be neglected and an elastic analysis is sufficient. Such an analysis is relatively simple and can usually exhibit the basic frequency behavior of the structure. Generalization from an elastic solution to include piezoelectric coupling is not always possible or simple analytically. An electrically forced vibration analysis including piezoelectric coupling is necessary when the electrical admittance (or impedance) of a device as a circuit element is wanted. Piezoelectric coupling also causes a stiffening effect that affects the resonance frequencies of the crystal. This needs to be considered in certain situations. Free vibration frequency analysis and electrically forced vibration analysis for admittance together provide a complete simulation of a crystal device. This paper includes results from both elastic and piezoelectric analyses.

## III. THICKNESS MODES

Consider a rectangular quartz plate of rotated Y-cut quartz as shown in Fig. 2. As a plate we assume the length $2a$ and the width $2c$ are much larger than the thickness $2b$. Pure thickness modes are modes that depend on the plate thickness coordinate $x_2$ and time only. They can only exist in plates unbounded in the $(x_1,x_3)$ plane ($a=c=\infty$ in Fig. 2) without edge effects. For thickness modes (2) reduces to two uncoupled groups:

$$c_{22}u_{2,22} + c_{24}u_{3,22} = \rho \ddot{u}_2,$$
$$c_{24}u_{2,22} + c_{44}u_{3,22} = \rho \ddot{u}_3, \tag{3}$$

$$c_{66}u_{1,22} + e_{26}\varphi_{,22} = \rho \ddot{u}_1,$$
$$e_{26}u_{1,22} - \varepsilon_{22}\varphi_{,22} = 0. \tag{4}$$

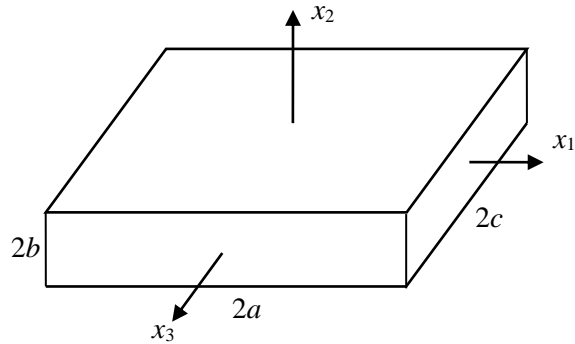

Fig. 2. An AT-cut quartz plate and coordinate system.

*A. Thickness-Shear Modes*

For time-harmonic motions, (3) and (4) become linear ordinary differential equations with constant coefficients. Therefore, problems related to thickness modes are always solvable mathematically. (3) is for the thickness-shear (TSh) mode in the $x_3$ direction described by $u_3(x_2,t)$ and the thickness-stretch (TSt) mode $u_2(x_2,t)$. These two modes are not coupled to the thickness electric field $\varphi(x_2,t)$. What is of more interest is the TSh mode in the $x_1$ direction with $u_1(x_2,t)$ governed by (4) which can be conveniently excited by a thickness electric field through the piezoelectric constant $e_{26}$. For the elastic case with the electric potential $\varphi=0$, simple solutions to (4)$_1$ for resonance frequencies and modes were obtained in [5] where the overtone frequencies are found to be integral multiples (harmonics) of the fundamental frequency. When piezoelectric coupling is included, the case when the two surfaces are electroded is more useful and the solution was given in [6] where the overtones are not harmonics. In resonator manufacturing and design, sometimes electrodes separated from the crystal plate with air gaps between them are used [7]. Distribution of $u_1$ along the plate thickness $x_2$ for static TSh deformation and the first few TSh modes are shown in Fig. 3. The fundamental mode with $n = 1$ is used most often in devices.

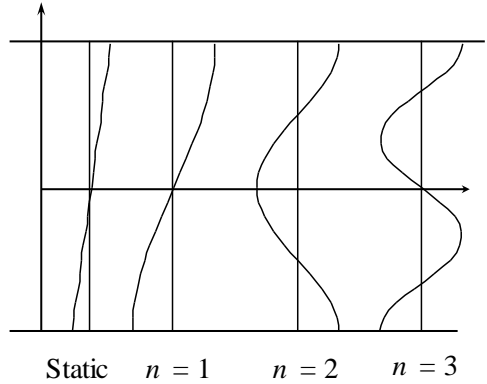

Fig. 3. Distribution of TSh deformation and modes along the plate thickness (the vertical axis).

When resonators are subjected to various environmental effects, e.g., a temperature change and stress, or there are additional surface structures like a mass layer (see Fig. 4 (a)), contact with a fluid (Fig. 4 (b)), particles (Fig. 4 (c)), and fibers (Fig. 4 (d)), the resonance frequencies of the TSh modes change. This effect provides the foundation for resonator-based acoustic wave sensors with broad applications. We discuss different types of these sensors separately below.

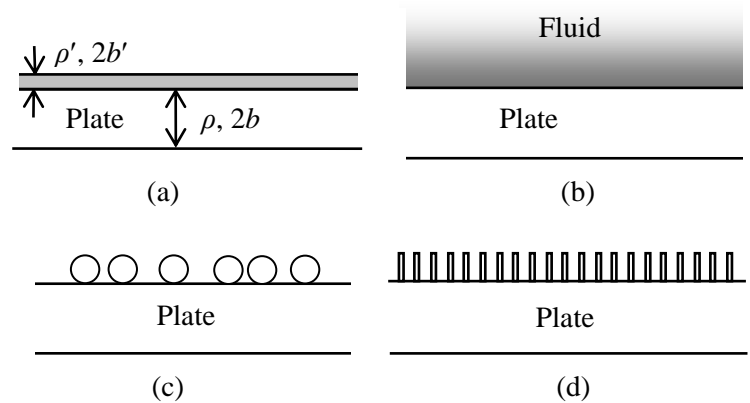

Fig. 4. Basic sensor structures.

## *B. Effects of Mass Layers*

When a thin layer of another material is attached to the surface of a crystal plate (Fig. 4 (a)), the TSh resonant frequencies of the plate become lower due to the inertia of the mass layer. This effect has been used to make quartz crystal microbalances (QCMs) for mass sensing. They are used for monitoring thin film deposition, and chemical as well as biological sensing. The sensitivity of a QCM is given by the well-known Sauerbrey equation [8] which was obtained by treating the mass layer effectively as a change of the plate thickness. Frequency shifts caused by two different mass layers on the top and bottom of a crystal plate representing the inertia of two electrodes of different thickness were obtained in [9] in an elastic analysis by solving the eigenvalue problem associated with the free vibration of a plate with mass layers. Piezoelectric solutions for TSh vibrations of plates with mass layers were obtained in [3] and [10]-[12]. [13] gives a solution for a plate of SC-cut quartz, a doubly-rotated cut. Mass sensitivity of a crystal plate or body can also be obtained from a perturbation integral [14]. In the above analyses of the mass effect, perfect bonding between the mass layer and the crystal plate was assumed, i.e., the displacement and traction are both continuous at the interface between the crystal and the mass layer. It was pointed out in [15] that imperfect bonding between the mass layer and the crystal plate often occurs, i.e., the motions of the mass layer and the plate surface may be different. A theoretical analysis on TSh vibration of a crystal plate with imperfectly bonded mass layers was presented in [16] where the interface was described by the so-called shear-slip model with interface elasticity which allows a discontinuity of the interface displacement. The result of [16] shows that an imperfectly bonded mass layer may lower or, somewhat counterintuitively, raise the resonance frequencies of resonators under certain conditions which had also been observed in experiments.

When the mass layers are not very thin compared with the crystal plate, in addition to its inertia, its shear modulus may also have a role. This effect was considered in [17]-[22] by applying the equations of elasticity or piezoelectricity to the mass layer with consideration of the variation of $u_1$ in the mass layer along its thickness. [17] and [20] are elastic analyses while the rest are piezoelectric. In particular, [19] considered a viscoelastic mass layer. [22] has some approximations. It considers the second-order effects of the variation of fields along the mass layer thickness by a power series expansion along the mass layer thickness with a truncation.

When there is interface elasticity or when the mass layers are thick and their shear modulus needs to be considered, an effective mass layer inertia which is frequency-dependent may be formally introduced. Then, depending on whether the effective inertia is positive or negative, the mass layer may lower or raise the resonant frequencies of the plate.

*C. Effects of Fluids*

A resonator in contact with a viscous fluid (Fig. 4 (b)) changes its resonant frequencies due to the inertia and viscosity of the fluid. This effect has been used to make fluid sensors for measuring fluid viscosity or density. For fluid sensor applications, the TSh mode of a plate without a normal displacement component at the plate surface is ideal. In this case, no pressure waves are generated into the fluid. The fluid produces a drag only on the plate surface, thereby lowering the frequencies. The classical result for frequency shifts cause by a fluid in contact with an elastic plate was given in [23]. It can also be obtained through a perturbation integral [24]. The problem of a fluid layer on a crystal plate under a separated electrode was studied in [25]. TSh modes in an AT-cut plate can also be excited using a lateral electric field through the piezoelectric constant $e_{36}$. A crystal plate in contact with a fluid under a lateral electric field was analyzed in [26]. The effects of simultaneous mass and liquid loading were studied in [27]. The effect of electrodes with finite thickness was examined in [28]. A relatively more complicated analysis on a multilayered structure was given in [29]. The effect of interface slip between a crystal plate and a fluid was investigated in [29, 30].

*D. Other Effects*

A resonator may be loaded with particles (Fig. 4 (c)). This often appears as an undesirable effect due to contamination. At the same time it may be considered as a way for measuring particle properties. While these particles are usually small compared to the thickness of the plate, in biosensing cells and certain other biological particles may not always be very small. TSh vibration of a plate carrying sparsely distributed, finite particles with a rotational degree of freedom and rotatory inertia was analyzed in [31] in which the particles can interact with the plate elastically. It was found that the particles may raise or lower the frequencies of the plate. Similarly, TSh vibrations of a plate with fibers (Fig. 4 (d)) have been explored for measuring fiber properties [32]. When the plate is in TSh vibration, the fibers are in flexural motions. The bending moments at the bottoms of the fibers effectively form a distribution of couples on the plate surface which cannot be accounted for using the theory of elasticity, and the more complicated couple stress theory of elasticity [33] is needed for a rigorous analysis [34]. If the TSt mode of the plate is used, the fibers are in extensional motions which is a simpler situation [35]. Quartz plates carrying beams immersed in fluids were analyzed in [36-39]. Quartz loaded with arrays of hemispheres were studied in [40, 41] to simulate cells on a QCM for biomedical applications. When a TSh resonator is subjected to an angular rate, its resonance frequencies may also be affected. This effect may be used to make gyroscopes or angular rate sensors. Thickness vibration of rotating crystal plates were studied in [42, 43] for this application. In addition to time-harmonic analyses for frequency behaviors, a dynamic analysis of a TSh resonator was performed in [44] to understand transient effects in a resonator such as its startup or turning off.

## IV. $x_3$-DEPENDENT MODES

Due to the finite size of real devices, edge effects are present and pure thickness modes cannot exist in finite plates. Therefore, in actual devices the operating TSh modes have slow in-plane variations in the $x_1$ and/or $x_3$ directions. These modes in finite plates are referred to as transversely varying TSh modes. In the case when the modes vary along one direction only, they are called straight-crested modes. For modes with in-plane variations along $x_3$ only, (2) splits into two uncoupled groups. The one of interest in this section is called the shear-horizontal (SH) or antiplane motion. It is described by

$$u_1 = u_1(x_2, x_3, t), \ u_2 = u_3 = 0, \ \varphi = \varphi(x_2, x_3, t), \tag{5}$$

and is governed by

$$c_{66}u_{1,22} + c_{55}u_{1,33} + 2c_{56}u_{1,23} + e_{26}\varphi_{,22} + e_{35}\varphi_{,33} + (e_{25}+e_{36})\varphi_{,23} = \rho\ddot{u}_1,$$
$$e_{26}u_{1,22} + e_{35}u_{1,33} + (e_{25}+e_{36})u_{1,23} - \varepsilon_{22}\varphi_{,22} - \varepsilon_{33}\varphi_{,33} - 2\varepsilon_{23}\varphi_{,23} = 0. \tag{6}$$

Compared to (2), (6) is much simpler and represents an important class of problems for which some results important to applications can be obtained theoretically.

### A. Waves in Unbounded Plates

SH motions include face-shear (FS) and thickness-twist (TT) modes. For FS waves, $u_1$ does not have nodal points (zeros) along the plate thickness. TT waves have nodal points along the plate thickness. Elastic and piezoelectric solutions for these waves were obtained in [45, 46] and [47], respectively. Their dispersion curves are shown in Fig. 5 where $\Omega$ is a dimensionless frequency and $Z$ is the dimensionless wave number in the $x_3$ direction. $m$ is related to the wave number in the plate thickness direction. The case of $m = 0$ is the FS wave. It is related to the straight line in Fig. 5 going through the origin and is nondispersive. The waves corresponding to the curves of $m > 0$ are the TT waves. They are dispersive. The finite intercepts of the curves with the $\Omega$ axis are the cutoff frequencies below which the wave number becomes purely imaginary and the corresponding waves cannot propagate.

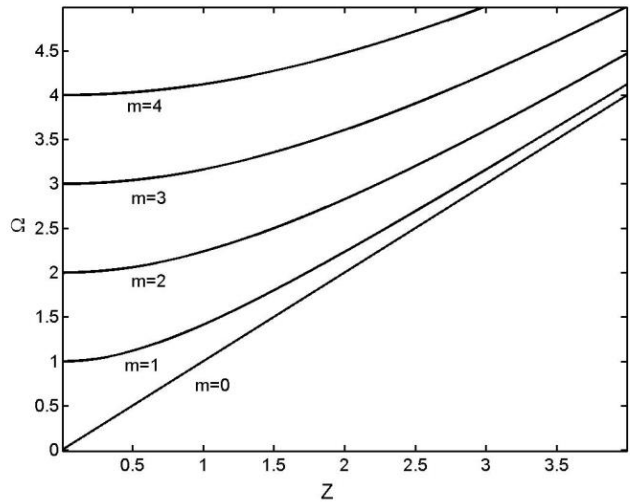

Fig. 5. Dispersion curves of FS ($m = 0$) and TT ($m > 0$) waves.

When a plate carries symmetric and asymmetric mass layers, the solutions for FS and TT waves were given in [46] and [48], respectively, using the equations of elasticity. The mass layer inertia lowers the dispersion curves in Fig. 5 slightly. When the mass layer is described by an elastic membrane governed by the plane-stress equations of elasticity, the dispersion curves for FS and TT waves were obtained in [49]. The effects of an imperfectly bonded mass layer with interface elasticity were studied in [50]. Thick mass layers with shear elasticity were treated in [51]. In all these cases [49]-[51], the interface elasticity or the mass layer in-plane and shear elasticity interact with the mass layer inertia to raise or lower the wave frequencies. Waves in a quartz plate on an elastic half-space were obtained in [52] which generalizes the well-know Love waves in elasticity propagating in an isotropic layer on an isotropic half-space. The effects of a viscous fluid on FS and TT waves were investigated in [53]. Waves in a plate with nonuniform material properties were studied in [54]. In [55], the case of a QCM carrying a porous mass layer was analyzed.

### B. Frequency Spectra of Finite Plates

For SH vibrations of a finite plate with thickness $2b$ and length $2c$, FS and TT modes are coupled by boundary or edge conditions. In resonator design, figures showing resonance frequencies versus the aspect ratio $c/b$ of the plate are called frequency spectra which is shown in Fig. 6 [56] for an AT-cut quartz plate. The curves in Fig. 6 are formed by data points without connecting them. They seem to automatically fall on some curves. Each data point represents the frequency of a mode under a given $c/b$. Corresponding to a particular value of $c/b$, there are infinitely many modes. A few can be seen in the frequency range shown in the figure. The curves may be classified into two families. The relatively steep family represents FS dominated modes. The relative gradual family is for TT dominated modes. When the flat parts of the TT family begin to bend near their ends or seem to be intersecting with the FS family, strong coupling between TT and FS mode occur which is undesirable in device operation and should be avoided. The usefulness of the frequency spectra is that it excludes a series of discrete values of $c/b$ which are associated with strong mode couplings.

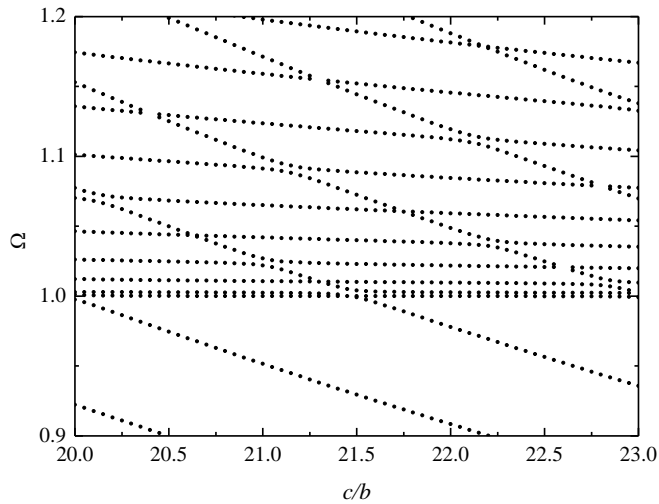

Fig. 6. Frequency spectra of coupled FS and TT modes near $\Omega = 1$.

### C. Energy Trapping and Bechmann's Number

The inertia of partial electrodes or mass layers can cause in-plane mode variations in a plate. This has important implications called energy trapping and Bechmann's number. To see these, consider a crystal plate with partial electrodes (mass layers) as shown in Fig. 7.

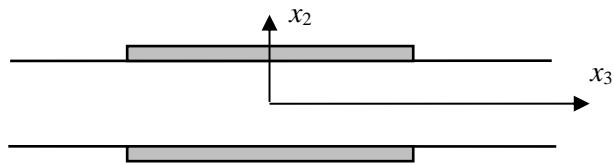

Fig. 7. An unbounded crystal plate with partial electrodes (trapped energy resonator).

In such a plate there may exist the so-called trapped modes that have oscillatory behavior in the electroded central region and decay exponentially out of the electrode edges. Thus the vibration energy is essentially trapped in the central region of the plate under the electrodes. This phenomenon is called energy trapping. Since the vibration essentially vanishes at a distance of a few times of the plate thickness away from the electrode edges, mounting of the crystal plate can be designed at such a distance without affecting the vibration of the plate. The reason why trapped modes may exist in a partially electroded plate can be qualitatively explained as follows. Figure 8 shows the dispersion curves of a TT wave in an infinite plate, either unelectroded or fully electroded. Consider long waves with a small wave number $\zeta$ which may be either real or pure imaginary. $\omega_1$ and $\omega_2$ are the cutoff frequencies of the wave. The inertia of the electrodes lowers the frequency. Therefore $\omega_2 < \omega_1$. If the partially electroded plate in Fig. 7 has resonances between $\omega_2$ and $\omega_1$, then the wave number in the electroded region is real but in the unelectroded region it is purely imaginary (see the intersections of the horizontal line with the dispersion curves). Therefore the modes have oscillatory behavior under the electrodes and decay exponentially away from the electrode edges in the unelectroded region.

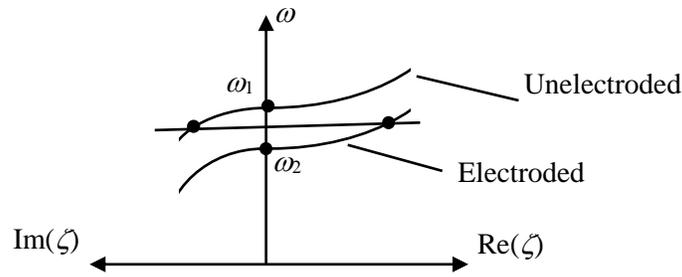

Fig. 8. Dispersion curves of TT waves in electroded and unelectroded plates.

There may be several trapped modes between $\omega_2$ and $\omega_1$, with different numbers of nodal points in the electroded region along the $x_3$ direction. The frequencies of these modes are closely distributed. In applications, usually only the first mode without a nodal point in the $x_3$ direction is desired. Bechmann observed that reduction of the electrode length eliminates the undesirable modes one by one until only the desired mode remains. Bechmann's number is defined as the ratio of the electrode length over the plate thickness below which only one trapped mode without a nodal point in the $x_3$ direction exists. Trapped TT modes were studied in [57, 58], with Bechmann's number given in [58].

*D. Trapped Energy Resonators and QCMs*

Even for the relatively simple case of elastic SH modes, exact solutions over finite domains can only be obtained for a plate with a pair of slightly tilted edges [59]. Since for AT-cut quartz plates we have $c_{55}$ = 68.81, $c_{56}$ = 2.53, $c_{66}$ = 29.01 $\times 10^9$ N/m$^2$ and $c_{56}$ is much smaller than $c_{55}$ and $c_{66}$, $c_{56}$ may often be neglected. This approximation makes a series of useful elasticity problems over rectangular domains solvable by trigonometric series. First consider the partially electroded rectangular resonator in Fig. 9 (a) [60]. One of its trapped modes is shown in Fig. 9 (b) for different electrode dimensions. Clearly, the vibration is mainly under the electrodes and decays away from their edges.

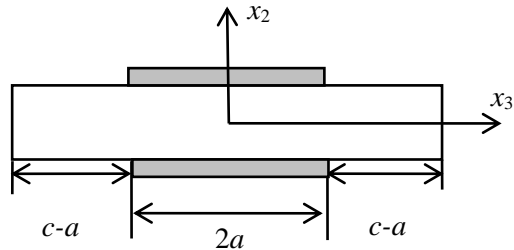

(a)

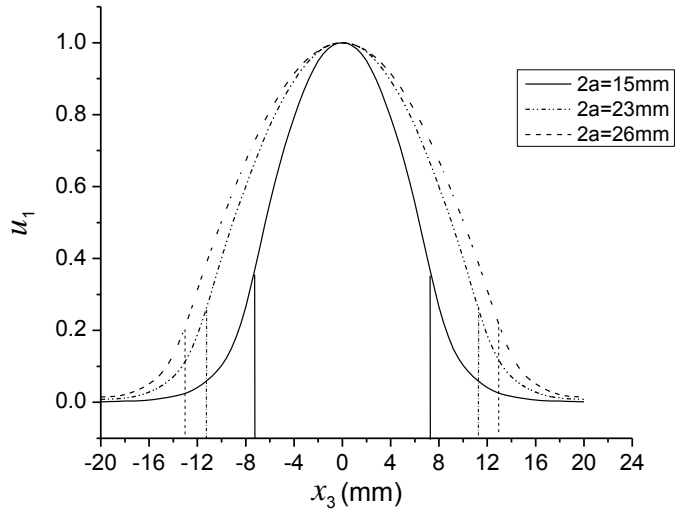

(b)

Fig. 9. (a) A finite plate with partial electrodes ($2c = 40$ mm). (b) Effects of electrode size on energy trapping of the first TT mode.

In resonator manufacturing, sometimes the top and bottom electrodes are slightly mismatched (see Fig. 10 (a)). This is a challenging problem in modeling. The only available theoretical result seems to be [61] which shows that the frequencies and modes are not very sensitive to electrode mismatch. However, it can be expected that the electrical admittance may be sensitive to mismatch. This problem requires a piezoelectric analysis and remains challenging. A QCM (Fig. 10 (b)) was analyzed in [62] and it was found that a single mass layer can also produce energy trapping. A plate with two pairs of electrodes as a filter (Fig. 10 (c)) and a plate with multiple mass layers as a QCM array (Fig. 10 (d)) were analyzed in [63] and [64], respectively. Their results show that the TT vibration may be trapped under some of the electrodes or mass layers, but not under the other electrodes. In general, longer or thicker mass layers tend to trap the vibration better. The solution for a mesa resonator (Fig. 10 (e)) with stepped thickness was obtained in [65] showing strong energy trapping. In [15], it was also pointed that the mass layer on a QCM is often nonuniform, in addition to that the mass layer may be imperfectly bonded to the plate. The case of a mass layer with a varying thickness and/or an imperfectly bonded mass layer was studied in [66]-[69]. In addition to the effects of the mass layer nonuniformity and imperfect bonding on frequency, the results of [66]-[69] show that the energy trapping is also sensitive to the mass layer thickness variation and the interface bonding.

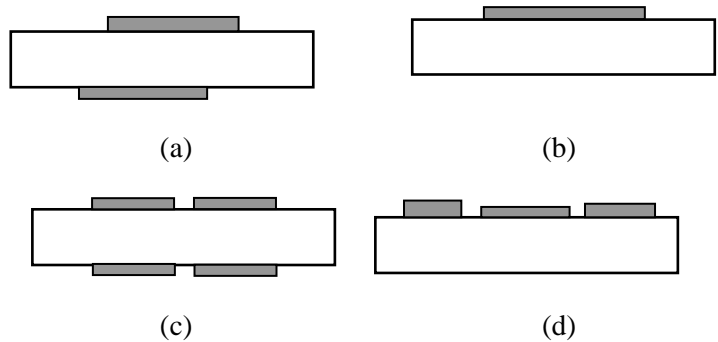

(a)            (b)

(c)            (d)

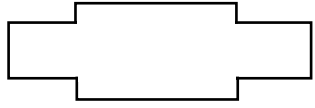

(e)

Fig. 10. Structures of resonators and mass sensors.

## V. $x_1$-DEPENDENT MODES

The governing equations for straight-crested modes with in-plane variations along the $x_1$ direction are obtained by setting the derivative with respect to $x_3$ in (2) to zero. For simplicity only the case of elasticity is shown below.

$$c_{11}u_{1,11} + (c_{12}+c_{66})u_{2,12} + (c_{14}+c_{56})u_{3,12} + c_{66}u_{1,22} = \rho\ddot{u}_1,$$
$$c_{56}u_{3,11} + (c_{66}+c_{12})u_{1,12} + c_{66}u_{2,11} + c_{22}u_{2,22} + c_{24}u_{3,22} = \rho\ddot{u}_2, \qquad (7)$$
$$c_{55}u_{3,11} + (c_{56}+c_{14})u_{1,12} + c_{56}u_{2,11} + c_{24}u_{2,22} + c_{44}u_{3,22} = \rho\ddot{u}_3,$$

which are clearly very much coupled.

### A. Propagating Waves

Waves propagating in an infinite plate can be formally obtained in general. For AT-cut plates they were given in [70] and were studied in detail in [71]-[73]. These waves can be separated into symmetric and antisymmetric waves according to whether $u_1$ is an even or an odd function of $x_2$ [71]. Figure 11 shows the dispersion curves of the antisymmetric waves [74]. The figure has many unconnected data points representing pairs of $(X, \Omega)$, the dimensionless wave number and frequency, satisfying the frequency equation. They form many curves representing different branches of the dispersion relations. For real and small values of $X$, when $\Omega$ increases from zero, the first three branches are the flexural wave dominated by $u_2$, the FS wave dominated by $u_3$, and the fundamental TSh wave dominated by $u_1$.

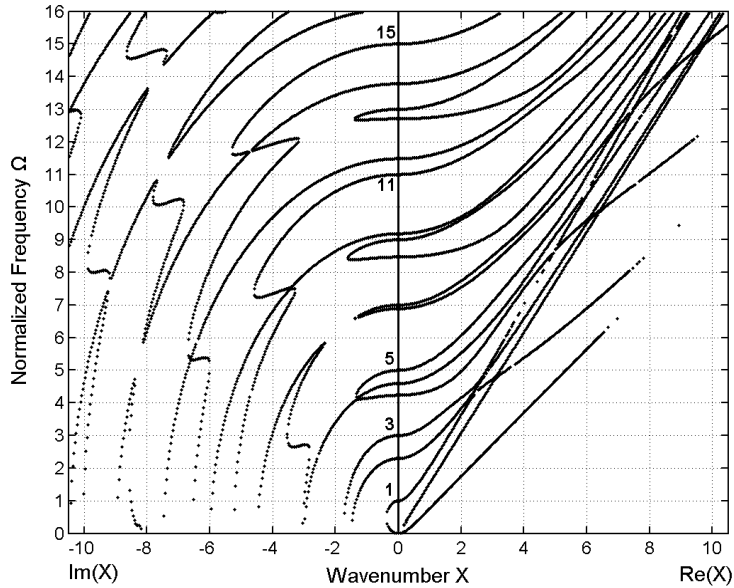

Fig. 11. Dispersion curves of antisymmetric waves in a quartz plate.

### B. Long Wavelength Limit

For long wavelength and low frequency, the wave number and the frequency approach zero. (7) degenerates to Cauchy's equations in (8) where the corresponding dispersion relations are [75]:

$$\gamma_{11} u_{1,11} = \rho \ddot{u}_1, \quad \Omega = X \sqrt{\frac{c_{11}}{c_{66}}},$$

$$-\frac{b^2}{3} \gamma_{11} u_{2,1111} = \rho \ddot{u}_2, \quad \Omega = \frac{\pi}{2} X^2 \sqrt{\frac{\gamma_{11}}{3 c_{66}}}, \quad (8)$$

$$\gamma_{55} u_{3,11} = \rho \ddot{u}_3, \quad \Omega = X \sqrt{\frac{\gamma_{55}}{c_{66}}}.$$

$(8)_1$, $(8)_2$, and $(8)_3$ are for extensional, flexural, and FS motions, respectively. The flexural and FS motions are antisymmetric according to the definition in the previous section and their dispersion curves can be seen in Fig. 11. The extensional motion is symmetric and its dispersion curve is not in Fig. 11. The dispersion curves for flexure and FS in (8) are local approximations of the corresponding dispersion curves in Fig. 11 for long waves and low frequencies.

In the region of long wavelengths and high frequencies, there are three families of modes in which the predominant displacement are TSh in the direction of the digonal axis of symmetry, TT (or TSh at right angles to the digonal axis), and TSt, respectively. Explicit formulas for the ordinates, slopes, and curvatures of the dispersion curves at infinite wavelength were given in [72]. These formulas are useful in the establishment of approximate, 2-D equations of motion that are solvable for high-frequency vibrations of bounded plates.

*C. Other Results*

In [76], propagation of straight-crested waves along the $x_1$ direction in a quartz plate with symmetric electrodes was studied. Dispersion relations and their asymptotic approximations for long waves were obtained from which Bechmann's number for a partially electroded plate was determined. Approximate solutions for coupled TSh, flexural, and FS vibration frequencies and modes in a finite plate bounded by $|x_1| < l$ were given in [77]. Waves in a doubly-rotated cut quartz plate exhibiting monoclinic symmetry were analyzed in [78]. A piezoelectric solution of straight-crested waves propagating in a plate of triclinic crystals was obtained in [79]. In [80], piezoelectric waves in a crystal plate with separated electrodes were obtained in general, with specific dispersion curves for AT-cut plates.

For AT-cut quartz plates, if the relatively small elastic constants $c_{14}$, $c_{24}$ and $c_{56}$ are neglected, (7) splits into two groups [81]. One is for the FS and TT waves described by $u_3$. The other is for coupled TSh and flexure and is very useful:

$$c_{11} u_{1,11} + c_{66} u_{1,22} + (c_{12} + c_{66}) u_{2,12} = \rho \ddot{u}_1,$$
$$(c_{66} + c_{12}) u_{1,12} + c_{66} u_{2,11} + c_{22} u_{2,22} = \rho \ddot{u}_2. \quad (9)$$

Dispersion curves for straight-crested waves governed by (9) were obtained in [81], along with solutions of finite plates using approximate boundary conditions in integral or resultant form. Coupled TSh and flexural vibrations represent the most important problem in resonator design. They are usually analyzed by approximate 2-D plate equations (see the review in [1]). (9) is not very complicated and more solutions of (9) useful in resonator design may be attempted.

## VI. MODES DEPENDING ON BOTH $x_1$ AND $x_3$

When the thickness modes have in-plane variations in both the $x_1$ and $x_3$ directions, the problem is 3-D. Theoretical solutions are impossible unless some approximations are made. For the most important TSh modes, the major displacement component is $u_1(x_1, x_2, x_3, t)$. In a somewhat crude approximation [82], $u_2$, $u_3$ and $\varphi$ are all neglected and the corresponding equations in $(2)_{2-4}$ are dropped. This results in the following single equation for $u_1$:

$$c_{11} u_{1,11} + c_{66} u_{1,22} + c_{55} u_{1,33} = \rho \ddot{u}_1. \quad (10)$$

(10) can be solved for a rectangular plate resonator [82, 83]. An analytical solution for a contoured resonator with a quadratically varying thickness for strong energy trapping was obtained in [84] in terms of Hermite polynomials. (10) was used to analyze a circular QCM in [85]. A solution to (10) in terms of trigonometric series for a partially electroded resonator (see Fig. 12 (a)) was given in [86], showing the energy trapping effect of the electrodes (Fig. 12 (b)).

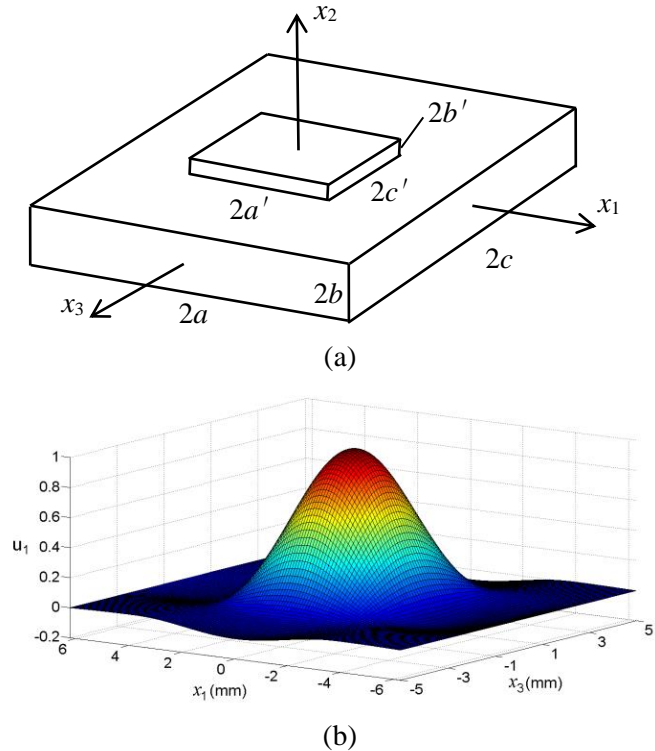

(a)

(b)

Fig. 12. A partially electroded resonator and energy trapping of the fundamental TSh mode.

## VII. CONCLUSION

The 3-D piezoelectric equations for quartz plates and various versions of their approximations as well as solutions of different models are reviewed. In spite of the complexity caused by material anisotropy, there still exists a significant amount of theoretical results. They provide fundamental understanding of the behaviors of resonators and acoustic wave sensors, and are the foundation for future theoretical research. They also serve as benchmarks for various approximation techniques, analytical or numerical. The relatively new crystals of the langasite family belong to the same crystal class as quartz but they have higher piezoelectric coupling. The results summarized in this review from piezoelectric analyses are applicable to langasite. Rotated Y-cut lithium niobate and lithium tantalate have the same elastic and dielectric matrices in (1) as quartz, but their piezoelectric matrix is complimentary to that of quartz in (1) in the sense that zero and nonzero elements are switched. Some of the theoretical studies on quartz plates may be helpful to the analysis of plates of lithium niobate and lithium tantalate or other crystals.

## ACKNOWLEDGEMENT

The author is grateful to Prof. Ji Wang of Ningbo University for providing Fig. 11.